\documentclass[prd, twocolumn, nofootinbib]{revtex4}

\usepackage{hyperref}
\usepackage{xcolor}
\usepackage{graphicx}
\usepackage{multirow}
\usepackage{amsmath}
\usepackage{rotating}

\newcommand{\vtheta}{\vec{\theta}}
\newcommand{\be}{\begin{equation}}
\newcommand{\ee}{\end{equation}}
\newcommand{\bel}[1]{\begin{equation}\label{#1}}
\newcommand{\ba}{\begin{eqnarray}}
\newcommand{\ea}{\end{eqnarray}}
\newcommand{\bal}[1]{\begin{eqnarray}\label{#1}}

\newcommand{\bside}{\begin{sideways}}
\newcommand{\eside}{\end{sideways}}

\newcommand{\rsol}{R_\odot}

\newcommand{\fns}[1]{\footnotesize{#1}}
\newcommand{\hpad}[2]{ \hspace{#2} #1 \hspace{#2} }

\newcommand{\jetangle}{\theta_J}
\newcommand{\Odds}{\mathcal{O}}
\newcommand{\Bayes}{\mathcal{B}}
\newcommand{\p}[1]{ p(\textrm{\footnotesize{#1}}) }
\newcommand{\pp}[2]{ \frac{ p(\textrm{\footnotesize{#1}}) }{ p(\textrm{\footnotesize{#2}}) } }
\newcommand{\snr}{\mty{SNR}}
\newcommand{\msm}[1]{\textrm{\footnotesize{#1}}}
\newcommand{\gam}{\textrm{\large{$\gamma$}}}

\newcommand{\xx}{\textrm{\msm{X}}}
\newcommand{\mty}[1]{\textrm{\tiny{#1}}}
\newcommand{\f}{\mathcal{F}}

\newcommand{\z}{$\zeta$}

\newcommand{\ergcms}{\textrm{ \fns{(erg cm}}^{\fns{-2}} \textrm{ \fns{s}}^{-1}\textrm{\fns{)}}}
\newcommand{\ergs}{\textrm{ \fns{(erg s}}^{-1}\textrm{\fns{)}}}
\newcommand{\ergcmshz}{\textrm{ \fns{(erg cm}}^{-2} \textrm{ \fns{s}}^{-1} \textrm{ \fns{Hz}}^{-1}\textrm{\fns{)}}}
\newcommand{\ergshz}{\textrm{ \fns{(erg s}}^{-1} \textrm{\fns{ Hz}}^{-1}\textrm{\fns{)}}}
\newcommand{\mr}[2]{\multirow{#1}{*}{#2}}

\newcommand{\mev}{$\msm{MeV}$}
\newcommand{\kev}{$\msm{KeV}$}

\newcommand{\nsbh}{NS-NS/NS-BH}

\newcommand{\fnm}[1]{\footnotemark[#1]}
\newcommand{\E}[1]{\hspace{-0.01in}\times\hspace{-0.02in}10^{#1}}

\begin{document}
\title[EM triggers for GW searches]{Electromagnetic transients as triggers in searches for gravitational waves from compact binary mergers}

\date{\today}

\author{Luke Zoltan Kelley\footnote{LKelley@cfa.harvard.edu}}
\affiliation{Harvard-Smithsonian Center for Astrophysics, 60 Garden Street, Cambridge, MA 02138}

\author{Ilya Mandel}
\affiliation{School of Physics and Astronomy, University of Birmingham, Edgbaston, B15 2TT, UK}

\author{Enrico Ramirez-Ruiz}
\affiliation{Department of Astronomy and Astrophysics, University of California, Santa Cruz, CA 95064}

\begin{abstract}
The detection of an electromagnetic transient which may originate from a binary neutron star merger can increase the probability that a given segment of data from the LIGO-Virgo ground-based gravitational-wave detector network contains a signal from a binary coalescence.  Additional information contained in the electromagnetic signal, such as the sky location or distance to the source, can help rule out false alarms, and thus lower the necessary threshold for a detection.  Here, we develop a framework for determining how much sensitivity is added to a gravitational-wave search by triggering on an electromagnetic transient.  We apply this framework to a variety of relevant electromagnetic transients, from short GRBs to signatures of  r-process heating to optical and radio orphan afterglows.  We compute the expected rates of multi-messenger observations in the Advanced detector era, and find that searches triggered on short GRBs --- with current high-energy instruments, such as Fermi --- and nucleosynthetic `kilonovae' --- with future optical surveys, like LSST --- can boost the number of multi-messenger detections by 15\% and 40\%, respectively, for a binary neutron star progenitor model.  Short GRB triggers offer precise merger timing, but suffer from detection rates decreased by beaming and the high \textit{a priori} probability that the source is outside the LIGO-Virgo sensitive volume.  Isotropic kilonovae, on the other hand, could be commonly observed  within the LIGO-Virgo sensitive volume with an instrument roughly an order of magnitude more sensitive than current optical surveys.  We propose that the most productive strategy for making multi-messenger gravitational-wave observations is using triggers from future deep, optical all-sky surveys, with characteristics comparable to LSST, which could make as many as ten such coincident observations a year.
\end{abstract}

\maketitle

\newpage

\section{Introduction}
\label{sec_intro}
The first generation of ground-based interferometric gravitational-wave (GW) detectors was successfully deployed during the last decade.  The Laser Interferometer Gravitational-Wave Observatory (LIGO\footnote{\url{http://www.ligo.caltech.edu}}), the Virgo detector\footnote{\url{http://www.virgo.infn.it}}, and the GEO-600 detector\footnote{\url{http://www.geo600.org}} have taken data that were analyzed to search for gravitational-wave signatures of compact binary mergers, short-lived transients, non-axisymmetric neutron stars, and stochastic gravitational-wave backgrounds \cite{InitLIGO, Virgo}.  No detections have been made so far, as searches have resulted in increasingly stringent upper limits on binary merger rates \citep[e.g.,][]{Collaboration:S5HighMass,S6lowmass}.  Currently, the network is undergoing upgrades to the ``advanced'' configuration, which will increase the detector sensitivities by a factor of $\sim10$, and thus the expected detection volume by $\sim10^3$ \cite{Smith:2009,AdvVirgo}.

Coalescences of compact-object binaries composed of neutron stars (NS) and black holes (BH) are expected to be among the most promising sources of gravitational waves for advanced LIGO and Virgo detectors.   The expected rates of local GW sources have been explored for over two decades \citep[e.g.,][]{nar91, phi91, kal04}.  Binary neutron stars (BNS) are predicted to merge at a rate of between 0.01 and 10 coalescences per Mpc$^3$ of comoving volume per Myr \cite{ratesdoc}.  The advanced GW detectors will be sensitive to BNS mergers to distances beyond $400$ Mpc for optimally oriented systems, with a typical range of $\sim 200$ Mpc.  At this sensitivity, even conservative predictions indicate that the first detections could be made soon after advanced LIGO/Virgo become operational in 2015-2016 \cite{ratesdoc}.

Coalescences involving neutron stars are also expected to produce a variety of electromagnetic transients.  First, and perhaps most notably, \nsbh~are promising candidates for the progenitors of short gamma-ray bursts (SGRBs) \cite[e.g.,][]{pac86, nar92}.  Unlike the long-duration majority of GRBs---associated with the final collapse of massive stars \cite[e.g.,][]{mac99}---SGRBs are believed to be jetted emission resulting from a small fraction of a solar mass rapidly accreting onto a stellar-mass black-hole \cite[e.g.,][]{lee07}.  In addition to gamma and \xx-ray emission, lower-energy observations have, on rare occasion, been associated with the `prompt' emission of long GRBs \cite[e.g.][]{ake99}, but so far not with short GRBs \cite{ukw11}.   Short and long GRB progenitors have also been associated with late afterglow  emission observed from the \xx-ray to the radio \cite[e.g.,][]{par97,fra97,cos97,bur06,sod06,geh09} on timescales between hours and weeks. In addition to the prompt and afterglow emissions associated with SGRBs, the neutron-rich tidal-ejecta---produced just prior to coalescence---have been proposed as a site of r-process element production \cite{lat76,fre99}, and associated optical emission (`kilonovae') on the timescale of about a day \cite{li98, kul05, ros05, met10, rob11, gor11, kor12}.  Throughout this paper, we quote results for the binary neutron star progenitor model, using the BNS GW detector horizon distance and inspiral rate; however, our methods are equally applicable to NS-BH binaries.

The prospect of a simultaneous detection of gravitational-wave and electromagnetic (EM) signatures from the same event would be particularly exciting.  Two channels of information from the same source would enable multi-messenger astronomy \citep[e.g.,][]{Bloom:2009}, probing the conversion of gravitational binding energy into electromagnetic radiation.  Additionally, the detection of a GW signal associated with a SGRB would be a unique and definitive determination of SGRBs' cryptic progenitors.   The observation of r-process `kilonovae', from a compact binary source confirmed by coincident GW detection, would permit a study of the associated densities and pressures through observations of nucleosynthetic products.  Electromagnetic observations can also yield redshift measurements, which would allow for alternative probes of cosmology when combined with distances extracted from GW signals \cite{Nissanke:2009}.

Rather than relying on serendipitous observations of EM and GW signatures from the same events, two alternative approaches to increase the detection rate have been proposed: (1) following up GW detection candidates with target-of-opportunity searches for EM counterparts, or (2) triggering searches of archival GW data based on observed EM transients (\cite{Mandel:2011} and references therein).

Approach (1) has been frequently discussed in the literature \cite[e.g.,][]{moh04,met11,cow11,nak11}, and recently, several follow-up searches for electromagnetic signatures associated with possible GW candidates have been carried out \cite{followupS6,SwiftS6}.  In the context of searches for binary mergers, approach (2) \cite[e.g.,][]{Stubbs:2008} has so far only been applied\footnote{High-energy neutrinos \cite{LIGOneutrino:2012} and soft-gamma repeaters / anomalous X-ray pulsars \cite[e.g.,][]{LIGOMagnetars:2011} have also been used as triggers in searches for GW signals.} to searches of GW data based on GRB triggers \cite{S5GRBLV,S6GRBLV} as proposed by, e.g.,~\citet{koc93} and \citet{fin99}.  The technical details of transient searches based on GRB triggers are discussed in \cite[e.g.,][]{HarryFairhurst:2011,Was:thesis}.  \citet{Nissanke:2013} and \citet{Dietz:2013} independently published analyses of related issues shortly after the preprint of the present paper appeared.

In this paper, we focus on approach (2) by performing a careful analysis of the improvement in sensitivity when information from an observed electromagnetic transient is used to trigger a GW search in archival LIGO/Virgo data.  Such information can boost the \textit{a priori} probability that a detectable GW signal exists in the data. EM triggers can further reduce the false alarm rate in GW searches by providing additional constraints on the sky location, inclination, or distance to the source.  We use a Bayesian framework (section \ref{sec:form}) to estimate the amount by which GW search thresholds can be reduced in EM-triggered searches relative to optimal, coherent all-sky searches.   We go beyond GRB triggers and consider a set of telescopes and EM surveys (section \ref{sec:em}) to identify classes of merger-associated electromagnetic transients that could increase detection rates. 

Using reasonable order of magnitude parameters (anticipating Sec.~\ref{sec:em}), we can motivate the characteristic scale of feasible improvements using simple Bayesian arguments (anticipating Sec.~\ref{sec:form}).
For simplicity, we will consider the case of a gravitational-wave search triggered by a SGRB observation.  The presence of an electromagnetic signal increases the {\it a priori} probability of a GW signal existing in a given segment of data at the sky location of the SGRB.  In particular, the change to the prior can be described as the product of the following quantities: the probability that the SGRB is associated with a binary merger (which we assume to be $\mathcal{F} \approx 1$); the probability that the merger occurred within the horizon of GW detectors, $\sim 400$ Mpc, despite the typical SGRB redshift being of order $0.7$ (see Table \ref{table_tran_obs}),
\begin{equation}
\frac{ V_\mty{GW} }{ V_\mty{EM} } \approx \left( \frac{ 400 \textrm{ Mpc} }{ 4000 \textrm{ Mpc} }\right)^3 \approx 10^{-3};
\end{equation}
and the increased expected rate of GW signals at a time close to the SGRB (typically chosen to be a $6$ second window around the SGRB trigger \cite{S6GRBLV}): if we expect $\sim 30$ events per year within the GW sensitive volume, the increase in the rate is
\begin{equation}
\mathcal{R}  \approx \frac{1 \, / \, 6 \textrm{ [sec/event]}} { 30 \textrm{ [events/year]} \, / \, 3\E{7} \textrm{ [sec/year]}} \approx 10^5.
\end{equation}
In addition to the improved prior, the presence of an EM signal localizes the analysis to a relatively small fraction of the sky (limited by the spatial resolution of GW detector networks),
\begin{equation}
\frac{ \Omega_\mty{GW} }{ \Omega_\textrm{sky} } \approx \frac{ 100 \textrm{ [sq.~deg]} }{ 40,000 \textrm{ [sq.~deg]} }  \approx 10^{-3},
\end{equation}
and with it, decrease the number of false alarms by the same fraction.  Combining these effects improves the overall a priori probability on the presence of a GW signal at the SGRB time and location by a factor of
\begin{equation}
\mathcal{P}_0 = \frac{ \mathcal{F}  \left( V_\mty{GW} / V_\mty{EM} \right)  \mathcal{R}}{ \left( \Omega_\mty{GW} / \Omega_\mty{sky}  \right) } \approx 10^5 .
\end{equation}

While this may seem like a large factor, it must be compared to a  typical likelihood threshold for detection.  A characteristic value for the LIGO network is a minimum signal-to-noise ratio (SNR) of 8 in at least two detectors (a `network' SNR $\approx 11.3$)---i.e. a ``likelihood threshold'' of $\mathcal{L}_0 = e^{ \mty{SNR}^2/2} \approx 6\E{27}$.  The SNR threshold in the presence of an EM counterpart required to achieve the same false alarm rate is
\begin{equation}
\textrm{SNR}_\mty{EM} = \sqrt{ 2 \ln{ \left( \frac{ \mathcal{L}_0 }{ \mathcal{P}_0} \right) } } \approx 10.2
\end{equation}
for a reduction of about $10\%$.  Thus, while it might naively seem as though triggered searches could drastically reduce the required SNR threshold by constraining the available parameter space, the benefits are much more modest.  The `glitchy' nature of GW data, fraught with large spikes of non-Gaussian noise, requires likelihood thresholds far out in the tails of the noise probability distribution - with minor improvements in the SNR threshold for large changes in the a priori probability of a detection.

Using the more detailed analysis that follows, we find that, for short GRB triggers (assuming these arise from compact-binary mergers), the threshold SNR reduction is a little less than 10\%---far more modest than the $\sim 50\%$ reduction predicted by \citet{koc93}.  This difference is due to a more realistic treatment of GW detector noise and GRB distribution and beaming angles in the present work, and is discussed in Sec.~\ref{sec:disc}.  The resulting increase in the total number of GW detections is marginal for triggered searches from any EM event, as only a fraction of mergers detectable with LIGO and Virgo are expected to have EM counterparts detectable with the concurrent generation of EM observatories.   The rate of multi-messenger observations, however, can be increased by as much as 30--40\% for optical, and 15\% for high-energy triggers.  While the benefits of triggering searches on short GRBs will be valuable for determining their progenitors, the most advantageous strategy for making multi-messenger observations is using r-process kilonovae detections from deep, optical all-sky surveys, such at LSST.

\section{Formalism}
\label{sec:form}
Standard, all-sky LIGO-Virgo searches for gravitational waves from coalescing compact binaries rely on matched filtering against a bank of template gravitational waveforms in order to extract signals from the noise \cite{findchirppaper}.  So far, most searches have not been fully coherent between detectors; instead, in coincident searches the best-fitting templates are found separately for each detector, and candidates are selected by searching for loud events in multiple detectors that are compatible in time and mass parameters \cite[e.g.,][]{Collaboration:S5HighMass,ihope}.  This approach is suboptimal when data from three or more detectors are available, since information about the relative phases and amplitudes in different detectors is not taken into account, and some candidates may not correspond to a self-consistent solution for extrinsic parameters, such as sky location, inclination, and distance to the source.  Fully or partially coherent searches have been developed \cite{Bose:2011,HarryFairhurst:2011}, but these are computationally expensive, and have not been regularly used except when an EM trigger is used to select the source location on the sky and to limit the time window, which makes a coherent search significantly less expensive.  

In this paper, we compare the improvements due to the presence of an electromagnetic transient trigger relative to a fully coherent blind all-sky search, in the anticipation that all-sky coherent searches will be computationally feasible when advanced detectors are operational.  Otherwise, if computational costs prevent coherent searches except when the sky location is known from the presence of an EM transient, the enhancements due to EM triggers will be even greater than we estimate.

\subsection{Bayes' Rule}
\label{sec_bayesrule}

The Bayesian formalism allows us to compare multiple hypotheses based on the given evidence and prior beliefs.  Consider an observed data set $d$ and a set of competing models $\{M_i | i = 1, 2, \ldots \}$ to explain that data, each with continuous parameters $\vtheta_i$.  Given a model and its parameters, the likelihood of observing the experimental data, $L(d|\vtheta_i, M_i)$, can be predicted.  Bayes' rule allows the posterior probability distribution function to be computed for a given set of parameters given the assumed model and the observed data:
	\be 
	p(\vtheta_i | d, M_i) = \frac{L(d|\vtheta_i, M_i) \, p(\vtheta_i|M_i)}{p(d|M_i)} \ ,
	\ee
where $p(\vtheta_i|M_i)$ denotes the a priori probability distribution of the model parameters before the data is taken into account. The denominator $p(d|M_i)$ is a constant determined by the requirement that posterior probability $p(\vtheta_i | d, M_i)$ must be normalized to one: 
	\bel{evidence}
  	p(d|M_i) = \int_{V_i} d\vtheta_i \, L(d|\vtheta_i, M_i) \, p(\vtheta_i|M_i) \ ,
	\ee
where $V_i$ is the parameter space volume in model $M_i$.  This value, known as the 'evidence', can be interpreted as the overall probability of generating the observed data set if model $M_i$ is correct.

Alternately, if we are interested in the posterior probability of a particular model given a set of data, $p(M_i | d)$, we can apply Bayes' rule as
	\be
	p(M_i|d) = \frac{p(d|M_i) \, p(M_i)}{p(d)}.
	\ee
Here $p(M_i)$ is the prior for model $M_i$, and $p(d)$ is another normalization constant defined discretely as
	\be
	p(d)=\sum_i p(d|M_i) \, p(M_i),
	\ee
under the assumption that all alternative models have been enumerated.  When comparing two alternative models, $M_1$ and $M_2$, this normalization cancels, so that the `odds ratio' between the models is just
	\be
	\Odds \equiv \pp{$M_1|d$}{$M_2|d$}  = \pp{$M_1$}{$M_2$} \, \pp{$d|M_1$}{$d|M_2$},
	\ee
where the first term is the ratio of prior probabilities of the models and the second term, equal to the ratio of their evidences, is known as the Bayes factor.

Using this formalism, we can compare a blind GW search with one triggered and constrained by an electromagnetic transient.

\subsection{Gravitational Wave Detection}
\label{sec_gwdetection}
In a given set of data, there is some probability that the data contains a (detectable) GW signal, $p(GW|d)$; and some probability that there is only noise, $p(N|d) = 1 - p(GW|d)$.  Specifically, we define $p(GW|d)$ as the probability that there is a GW signal in the data ending in a time interval of duration $\tau_\mty{GW}$ (ms) from a binary inspiral within a distance $\delta$ (Mpc)\footnote{We use $\tau_\mty{GW}=100$ ms, corresponding to a typical time window of a coherent Bayesian analysis; and $\delta=1000$ Mpc, which is sufficiently large to ensure that all binary neutron star mergers detectable as GWs fall within this volume.  The exact values are not important -- they are only relevant for making the Monte Carlo simulation described in section \ref{sec_implementation} more efficient.}  The odds ratio for the data to contain such a signal is
	\be \label{eq_odds}
	\Odds \equiv \pp{GW$|$d}{N$|$d} = \pp{GW}{N} \, \pp{d$|$GW}{d$|$N}.
	\ee
The factor $\frac{p(GW)}{p(N)}$ is the prior probability of the data containing a signal, regardless of the data collected, primarily determined by the expected merger rate of compact binaries within the distance $\delta$.  	The Bayes factor, $\Bayes \equiv \frac{p(d|GW)}{p(d|N)}$, on the other hand, is a purely \textit{a posteriori} measure of the confidence in having observed a GW signal:
	\be \label{eq_bayes_factor}
	\Bayes = \pp{d$|$GW}{d$|$N} =\frac{\int  L(\vec{\theta}|GW) p(d|\vec{\theta}) d\vec{\theta}}{p(d|N)}.
	\ee
Under the assumption of stationary, Gaussian noise, the Bayes factor can be approximated as a function of the signal-to-noise ratio (SNR),
	\be  \label{eq_snr}
	\Bayes \propto \eta e^{ \frac{1}{2} ( \snr )^2 },
	\ee
where $\eta$ is the fraction of the prior volume of the parameter space to which the signal's parameters can be constrained\footnote{The fraction of prior volume in which the likelihood is significant, $\eta$, in turn depends on the SNR: higher SNRs yield better parameter constraints and smaller $\eta$.  We neglect this dependence here.} (see appendix of \citep{Veitch:2012}, where an exponential is missing from Eq.~(A.5)).  Therefore, reducing the prior volume by a particular factor {\it in the region where the likelihood is negligible} is equivalent to an increase in the evidence integral by the same factor.  

Combining Eqs. \ref{eq_odds} \& \ref{eq_snr}, the odds ratio can be expressed as,
	\be  \label{eq_odds_snr}
	\Odds \propto \pp{GW}{N} \cdot \eta \cdot e^{ \frac{1}{2} ( \snr )^2 }.
	\ee
The confidence in detecting a signal ($\Odds$) is thus tied to the observed SNR.  In practice, a `detection' is defined by a threshold on the odds ratio such that the false-alarm rate is reduced to some (small) predetermined value (e.g.,~$< 10^{-2} \textrm{ yr}^{-1}$).  

Although the odds ratio as written in Eq.~(\ref{eq_odds_snr}) corresponds to the assumption of stationary and Gaussian noise, we know from experience that LIGO and Virgo noise has non-stationary, non-Gaussian artifacts (`glitches').  A confident detection may require an SNR threshold of $\sim 8$ in at least two detectors, correspond to a network SNR of $11.3$  \citep[e.g.,][]{S6lowmass}.  Therefore, in our analysis we choose to use the artificially large odds ratio threshold of $6\E{15}$, based on $\rho=11.3$, $\eta=10^{-5}$ (see table \ref{table_eta}) and $p(GW) \approx \pp{GW}{N} =10^{-7}$ (using the pessimistic merger rate from from \citet{ratesdoc}\footnote{For a binary neutron star merger rate of $0.01 \textrm{ Mpc}^{-3} \textrm{ Myr}^{-1}$~\cite{ratesdoc}, with $\tau_\mty{GW} = 100$ ms  and $\delta = 1000$ Mpc, $p(GW)=10^{-7}$.}) in order to empirically account for the data quality in a conservative manner.  The same odds ratio threshold is assumed to hold for GW-only candidates and for those with electromagnetic triggers, $\Odds_\mty{GW} = \Odds_\mty{EM} = 6\E{15}$.  

Our analysis is fairly insensitive to the choice of the odds ratio threshold---or, alternatively, the choice of the priors---due to the large uncertainties in other parameters and the super-exponential dependence of the odds ratio on the SNR.  However, we note that a lower odds ratio threshold---as would result from more stationary and gaussian detector noise---would make the corresponding SNR threshold more sensitive to constrains on the prior and parameter spaces from EM triggers.

\subsection{Electromagnetic Counterparts to GW Mergers}
\label{sec_emcounterparts}
Additional evidence for the presence of a gravitational wave signal decreases the required signal-to-noise ratio for a given overall detection confidence (odds-ratio).  Because the SNR is inversely proportional to the source distance for a given type of inspiral event, increasing the effective SNR amounts to increasing the detector horizon by the same factor, and the accessible detector volume by its cube.  Thus, even a small improvement to the SNR can have a large effect on the expected detection rate.  We can compute the fractional increase in sensitivity, i.e., the fraction by which the SNR threshold can be lowered, for a given event, when an EM transient trigger has been observed (denoted by the condition $|EM$), by rearranging Eq. (\ref{eq_odds_snr}):
	\be
	\label{eq_snr_ratio}
	\zeta	 \equiv 	 \frac{ \msm{SNR} }{ \msm{SNR}_\mty{$|$EM} }		
	= \left[  		\frac{ \textrm{ ln}\left( \Odds_\mty{GW} \cdot \left[ \pp{GW}{N} \cdot \eta_\mty{GW} \right]^{-1} \right) }
							{ \textrm{ ln}\left( \Odds_\mty{EM} \cdot \left[ \pp{GW$|$EM}{N$|$EM} \cdot \eta_\mty{EM} \right]^{-1} \right) }		\right]^\frac{1}{2}.	
	\ee
If the GW prior is significantly increased by the presence of an electromagnetic counterpart, i.e., $\pp{GW$|$EM}{N$|$EM} \gg \pp{GW}{N}$, then $\zeta > 1.0$ and the required $\Bayes$ (i.e.,~SNR threshold) is decreased relative to an analysis of the GW-data alone.  The presence of an EM signature also affects $\zeta$ through the parameter space term ($\eta$).  Due to the higher spatial resolution of photonic observations \cite[see, e.g.,][]{fai09}, only a portion of the total GW parameter space will be compatible with an EM signature.  The effects of an EM counterpart on the GW-prior and parameter-space terms are examined in the following sections.  Note that the division of information from an EM transient into two distinct terms, prior and parameter-space, is purely conceptual.  Constraints present in both terms have the same effect of ruling out false alarms and improving the chances of detecting a GW signal.

\subsubsection{GW Prior}
\label{sec_gw_prior}
The prior probability of a data-segment containing only noise (i.e.~no detectable GW signals) can be expressed as $\p{N} = 1 - \p{GW}$ and $\p{N$|$EM} = 1 - \p{GW$|$EM}.$\footnote{It is also possible that, given an EM signal, there could be a chance, \textit{un}associated GW-signal detectable in a blind search; we do not consider this possibility here.}  The prior for a detectable GW is given by the expected merger-rate density $\mathcal{R}$, multiplied by the detector volume and time-duration of the data segment: 
	\be
	\label{eq_gwprior}
	\p{GW} =  \mathcal{R}  \frac{4\pi}{3}\delta^3 \, \tau_\mty{GW},
	\ee
where $\tau_\mty{GW}$, which is in principle arbitrary, is chosen to be small enough such that $\p{GW} \ll 1$, and also that $\tau_\mty{GW} < \Delta t$ in Eq.~\ref{eq_priorspace}.  The exact choice of $\tau_\mty{GW}$ does not impact our results as it formally cancels in the $\Odds$ and $\p{GW}$ terms in Eq.~\ref{eq_snr_ratio}, and is only used for convenience in Monte Carlo simulations.

In the case of an electromagnetic counterpart, the GW prior must take into account the probability that the EM signature was produced by a merging binary ($\f$), as well as the probabilities that the merger took place within the time interval  $\tau_\mty{GW}$ and within the distance $\delta$.  More precisely, the probability of having a coalescence given a particular type of electromagnetic signature is $\f$.  In our analysis, we optimistically assume that $\f$ is unity; however, even assuming $\f = 0.1$ would decrease the multi-messenger detection rates by less than 5\% for some of the transients.  In general, a triggered search yields benefits over a blind all-sky search as long as the product of the prior probability for a GW signal and reduced parameter volume $\eta$ (discussed in the next subsection) is greater given an EM transient trigger than without one.   This condition yields a threshold for $\f$, which, given a particular EM transient, determines whether or not triggering will be beneficial.

If we denote the temporal localization of the merger based on the electromagnetic signature by $\Delta t$, and the electromagnetically accessible volume by $V_\mty{EM}$, the probability that there was a binary merger within the time $\tau_\mty{GW}$ and distance $\delta$ given the EM transient observation is,
	\be
	\label{eq_priorspace}
	\p{GW$|$EM} = \f \cdot \frac{ \tau_\mty{GW} }{ \Delta t} \cdot \min\left(\frac{(4/3)\pi \delta^3}{V_\mty{EM}} \, , \, 1.0 \right).
	\ee
For a SGRB, $ \tau_\mty{GW} / \Delta t$ would be relatively large---as SGRBs are expected to quickly follow the merger \cite{nak11}.  Orphan afterglows (those observed without prompt SGRB signals), on the other hand, would be expected on much larger timescales (days to months)---leading to a correspondingly smaller prior.  In general, higher telescopic precision better constrains the parameter space---increasing $\zeta$ (\ref{sec_parspace}).  For example, if the EM signal has a distance determined to be within the GW-horizon, the volume term in Eq.~(\ref{eq_priorspace}) goes to unity.  If, however, there is no distance measurement (i.e., no redshift), higher sensitivity increases the accessible volume ($V_\mty{EM}$)---and lowers $\zeta$, because \textit{a priori} it becomes less likely that these transients come from within the sensitive volume of GW detectors.

\subsubsection{Parameter Space}
\label{sec_parspace}
Given a coincident detection of a GW signal and an EM signal from the same source, the EM uncertainty in the sky location (and perhaps distance and inclination) will be significantly less than that for the GW signal alone. Therefore, a more restrictive prior can be used for GW analyses based on EM triggers.  Consequently, the fraction of the prior volume for the parameter space of a GW observation, $\eta$, will be enhanced by the EM observation.  We can decompose $\eta$ into distance-inclination (partially degenerate for GW observations \cite{Veitch:2012}) and angular terms,
	\be
	\label{eq_parameterspace}
	\eta = \eta_d \cdot \eta_\phi.
	\ee
We can then compare the wide prior volume which must be considered in blind all-sky gravitational-wave searches and the restricted prior volume when an electromagnetic counterpart has been detected.

Without an EM signature, the angular term may be on the order of $\eta_\phi \msm{(GW)} \equiv \Omega_\mty{GW} / \Omega_\msm{sky} \approx 10^{-3}$, where $\Omega_\mty{GW}$ is the GW detector angular uncertainty---on the order of tens of square degrees \cite{fai09,Veitch:2012}; while the distance-inclination fractional parameter space might be an order of magnitude less constrained, i.e.~$\eta_d \msm{(GW)} \approx 0.01$.  With the presence of an electromagnetic counterpart, the angular localization of an EM signature (generally arcseconds to arcminutes) is always more accurate than that of a GW antenna: $\Omega_\mty{EM} < \Omega_\mty{GW}$, and thus $\eta_\phi \msm{(GW$|$EM)} = 1$.  Similarly, if the EM transient yields a distance determination (i.e., a spectroscopic redshift from the signal itself, or an associable host galaxy), $\eta_d \msm{(GW$|$EM)} = 1.$\footnote{Strictly speaking, even when orphan afterglows enable redshift determination, the inclination of the source may be poorly constrained, while observations of SGRB prompt emission will constrain the source inclination to within the jet opening angle.}  If a distance determination is impossible, the distance parameter space reduces to that of a GW-signal alone.  These values are summarized in Table \ref{table_eta}.

\begin{table}[h] 
\renewcommand\arraystretch{1.2}
\centering      
\begin{tabular}{ c @{\hspace{-2em}} c @{\hspace{-1em}} l @{\hspace{-0em}} r @{\hspace{-1em}} c}
							& \hpad{\large{$\eta_\phi$}}{0.3in}		& \multicolumn{2}{c}{\large{$\eta_d$}}		& \hpad{\large{$\eta$}}{0.4in}		\\ \hline 
\hpad{GW}{0.1in}				&  $10^{-3}$						& \multicolumn{2}{c}{$10^{-2}$}			& $10^{-5}$					\\ \hline
\multirow{2}{*}{GW$|$EM}			& \multirow{2}{*}{1}					& with redshift:		& \hpad{1}{0.1in}		& 1.0							\\ \cline{3-5} 
							&								& without:			& $10^{-2}$			& $10^{-2}$					\\ \hline
\end{tabular} 
\caption[Fractional Parameter Space for GW and GW-EM Detections.] {The fraction of the prior volume of the parameter space to which the source can be constrained for detections of GWs alone, and GWs with an EM counterpart is decomposed into angular and distance-inclination components.  An electromagnetic component will always greatly improve the angular localization, while the distance will only be better-constrained with redshift determination.}
\label{table_eta}
\end{table}

\subsection{Implementation}
\label{sec_implementation}
We use a Monte Carlo simulation to implement this formalism in determining plausible detection properties.  For every telescope and transient combination, binary merger events are distributed in space according to a fixed merger rate density per unit source time per unit comoving volume.  We use the default rate of 1 Mpc$^{-3}$ Myr$^{-1}$ from \citet{ratesdoc}.  The source sky location $(\theta, \phi)$, inclination $\iota$, and polarization $\psi$ are chosen isotropically.  Lightcurves (section \ref{sec:em}) are interpolated to the relevant viewing angle, and detectability is determined.  The `observed' properties of the EM transient are then fed into a calculation of $\zeta$ according to Eqs.~\ref{eq_snr_ratio}, \ref{eq_gwprior}, \ref{eq_priorspace} and Table~\ref{table_eta}.  We model redshift as being determined by a spectroscopic follow-up if the time-over-threshold (ToT) of the observable EM signal is longer than one day {\it and} the EM signal undergoes an e-folding in amplitude during this time.  These requirements are imposed to roughly account for the subset of detected signals which warrant followup, and the time needed to obtain it.  How accurately the merger-time can be determined based on the EM signal depends on how well sampled, and how well modeled the light-curve is.  As a conservative approximation, we assume the merger-time is only as well constrained as the typical time-at-maximum (TaM) of that signal.  The SNR of the GW signal alone is calculated for a single detector based on the source distance and sky location and inclination angles according to the equation
	\bel{SNR}
	\textrm{SNR} = 2.0 \, (1 + z)^{5/6} \left(\frac{d}{d_H}\right)^{-1} \Theta(\rm{angles}),
	\ee
for a source at redshift $z$ and luminosity distance $d$. The detector horizon distance $d_H = 445$ Mpc from \citet{ratesdoc} is defined as the distance at which  the SNR for an optimally located and oriented source equals $8.0$.  The factor of $(1+z)^{5/6}$ is based on the scaling of the waveform amplitude with the (redshifted) chirp mass, and is accurate for low-mass systems whose SNR is limited by the bandwidth of the detector rather than the ending frequency of the GW signal.  The angular dependence $\Theta(\rm{angles})$ \cite[e.g.,][]{fin93,tay12}, is given by
	\be
	\Theta \equiv 2 \left[ F_+^2 \left(1 + \cos^2 \iota \right)^2 + 4 F_\times^2 \cos^2 \iota \right] ^{1/2},
	\ee
with antenna pattern projections,
	\begin{align}
	F_+ & \equiv \frac{1}{2} \left(1 + \cos^2 \theta \right) \cos 2\phi \cos 2\psi - \cos \theta \sin 2 \phi \sin 2 \psi, \nonumber \\
	F_\times & \equiv \frac{1}{2} \left(1 + \cos^2 \theta \right) \cos 2\phi \sin 2\psi + \cos \theta \sin 2 \phi \cos 2 \psi.
	\end{align}
Cosmological distance measures are converted using WMAP-7 parameters \cite{wmap7}, included in Table~\ref{table_params}.  GW detectability is then determined by the threshold $\textrm{SNR} \geq 8$ in the absence of an EM trigger, and $\textrm{SNR} \geq 8/\zeta$ when an EM transient is observed.  The conversion between merger rates in source and observer times is done probabilistically --- an event in a year of source time has a probability of $(1+z)^{-1}$ of being detected in the corresponding observer-year.  Finally, bootstrapping from the subset of simulated detections, is used to estimate parameter variance.


\section{Electromagnetic transients} \label{sec:em}

Gravitational wave sources in the LIGO-Virgo sensitive band ($\sim$100 Hz) are dominated by neutron-star and stellar-mass black-hole binaries in the final seconds before coalescence.  Out of the three permutations of source types, roughly 10 binary neutron-star systems have been observed in our Galaxy \citep[e.g.,][]{wol89, wol91, kiz10}.  Four of those systems have sufficiently small orbital separations---$a \lesssim 5 \, \rsol$, or $P \lesssim 0.5 \textrm{ day}$---that gravitational radiation will merge the system within a Hubble time.  Exotic formation channels are required to produce such systems starting with an initial binary of two massive stars, including two core-collapse supernovae and most likely a phase of common-envelope evolution \citep[e.g.,][]{por98,fry99,Kim:2006,kal07,lee07}.  Electromagnetic transients associated with GW mergers most likely require the presence of at least one NS.  If the binary reaches the Roche limit, the NS will be tidally disrupted---ejecting a few hundredths of a solar mass in one or two tidal tails for NS-BH and NS-NS systems, respectively \citep[e.g.,][]{lat74,ros99,ros05,lee07,rob11}.\footnote{Several intriguing scenarios for \textit{pre}-merger electromagnetic signatures have been proposed \cite[e.g.,][]{HansenLyutikov:2001,Lehner:2011,Tsang:2012}, but are not considered here.}

There is a growing consensus that the expansion of neutron-rich material and r-process powered nuclear heating can act as an effective energy reservoir to power fast optical transients \cite{met10,rob11}, although the precise peak timescale and temperature is dependent on the opacity of r-process nuclei, which is currently not well constrained \cite{kase13, barn13}. The most efficient conversion of radioactive energy to radiation is provided by those isotopes with a decay timescale comparable to the radiative diffusion time through the ejecta.

The high angular momentum bulk of material from the merger will form a transient disk around the existing or newly formed black hole.  Cooling is neutrino dominated and pair production and/or a relativistic MHD wind can lead to a jetted outflow with $\Gamma \approx 100$ \citep{lee07}, while the disk is rapidly consumed on an accretion timescale \cite{pop99,nar01,lee04, set04, lee05,met08,lee09}, on the order of a second.  Because the emitting region must be several powers of ten larger than the compact binary that acts as trigger, there is a further physical requirement: the original beamed,  relativistic outflow would, after expansion, be transformed into bulk kinetic energy. This energy cannot be efficiently radiated as gamma rays unless it is re-randomized, which requires relativistic shocks. The gamma-rays we receive come from only the material whose motion is directed within $1/\Gamma$ of our line of sight---which must lie within the jet angle $\theta_j$.  
At observer times of more than about a week, the blast wave has been decelerated to a moderate Lorentz factor, irrespective of its initial value.  In this `afterglow' phase, beaming and aberration effects are less extreme, emission is observable from a wide range of angles, and is thus sensitive to the ejecta geometry \cite{rho99}.  The minimum random Lorentz factor of protons going through the decelerating shock is expected to be comparable to the bulk Lorentz factor, while that of the electrons may exceed this by a factor of up to the ratio of the proton to the electron mass. The energy of the particles can be further boosted by diffusive shock acceleration as particles scatter across the shock interface repeatedly, acquiring a power law distribution $N(\gamma)\propto \gamma^{-p}$, where $p\sim 2-3$. In the presence of turbulent magnetic fields built up behind the shocks, the electrons are expected to produce a synchrotron power-law radiation spectrum.

For an approximately smooth distribution of external matter, the bulk Lorentz factor decreases inversely with time and, as a consequence, the minimum accelerated electron random Lorentz factor and the amplified magnetic field also decrease. This implies that the spectrum softens in time, leading to late optical and radio afterglow emission \cite{coll12}.  As the bulk material decelerates, afterglow emission peaks at progressively lower frequencies.  Off-axis observers see a rising light-curve reaching a peak when the Lorentz factor drops to $\Gamma(t) \approx 1/\theta_\msm{obs}$, followed by a power-law decrease in luminosity asymptotically approaching the on-axis lightcurve \cite{der00,gra02,ram05,van11}.   In modeling prompt and afterglow emission, the jet angle can be inferred by matching both the observed SGRB rate (assuming BNS progenitors), and the observed afterglow luminosities.  Based on these requirements, we use $\theta_j \approx 0.2$, consistent with jet breaks observed in GRB afterglows \cite[e.g.,][and references therein]{fon12}.

\subsection{Instruments and Surveys}
To explore plausible detection scenarios, we use the parameters of several telescopes and surveys across the electromagnetic spectrum.  The values used, while consistent with each instrument's characteristics, should be taken as representative in an order-of-magnitude sense.  Some of the instruments addressed have temporary or as-of-yet undecided strategies and time-allocations which will decrease their overall time-sky coverage.

In the optical, we explore both the Palomar Transient Factory \cite{rau09} (PTF) and Large Synoptic Survey Telescope \cite{ive08} (LSST), which are designed for deep, fast-transient surveys of large fractions of the sky.  Numerous other optical telescopes are in development or already exist (e.g.,~Pan-STARRS \cite{kai02}), but have been omitted due to their parametric similarity to PTF/LSST, or their focus on longer-cadence observations (e.g.,~SkyMapper \cite{kel07}).  In the radio we examine three surveys---Apertif \cite{oos10}, ASKAP \cite{joh09}, and LOFAR \cite{fen06}.  Apertif is currently taking proposals for survey strategies; out of a range of options, we choose (arbitrarily) a very narrow, very deep survey to juxtapose with the shallower, wider ASKAP survey.  For high-energy observations, we consider the Swift satellite's Burst Alert Telescope (BAT) \cite{bar05} in the \xx-ray regime (15 -- 150 \kev), and the Fermi Gamma-ray Burst Monitor (GBM) \cite{mee09}  which extends into gamma-rays (8 \kev~-- 40 \mev).  The parameters used for each instrument and survey are presented in Table~\ref{table_instr}.

For detector horizons on the scale of Gpc and larger, cosmological effects become important.  We assume that the merger rate \cite{ratesdoc} is constant in comoving volume until the star-formation-peak at $z \approx 1.5$ \cite{mad96,mad98} (luminosity distance $d_L \approx 11\textrm{ Gpc})$, and is negligible earlier \cite{kel10}.  Detectability is calculated in temporally-redshifted luminosity space, but spectra are assumed to be approximately constant between the source and detector frames (i.e.~`K corrections' are not considered).

\begin{table*}[ht] 
\renewcommand\arraystretch{0.4}
\centering      
\begin{tabular}{ c c c c c c }
	Project 					& Band					& \hspace{0.8in}Sensitivity\hspace{0.8in}		& FoV (Survey)		& Cad.(d)					& Ref. \vspace{0.05in}		\\ \hline
	\mr{2}{Swift \fns{(BAT)} }		& \xx-ray 					& \mr{2}{$10^{-8} \ergcms$}				& 4,600 			& \mr{2}{2}	 			& \mr{2}{ \citep{cus10,bar05} } 	\\ 
							& \fns{(15 - 150 \kev)} 		&									& (40,000)		&						& \vspace{0.04in} 			\\  \hline 
	\mr{2}{Fermi \fns{(GBM)} }		& \gam-Ray				& \mr{2}{$10^{-6} \ergcms$ \fnm{1}}			& 31,200			& \mr{2}{1}				&  \mr{2}{ \citep{mee09} }           	\\  
							& \fns{(8 KeV - 40 MeV)}		&									& (40,000)		&						& \vspace{0.04in}			\\ \hline 
	\mr{2}{LSST}				& Optical					& 24.5								& 9.6		 		& \mr{2}{3} 				& \mr{2}{ \citep{ive08} } 		\\ 
							& \fns{(r: 550 - 700 nm)}		& $5.8\E{-30} \ergcmshz$					& (10,000)  		&						& \vspace{0.04in}			\\ \hline
	\mr{2}{PTF} 				& Optical					& 21.0								& 7.9 			& \mr{2}{5} 				& \mr{2}{ \citep{rau09,law09} } 	\\ 
							& \fns{(r)}					& $1.4\E{-28} \ergcmshz$					& (8000) 			&						& \vspace{0.04in}			\\ \hline
	\mr{2}{Apertif} 				& Radio 					& $0.1 \, \mu \textrm{Jy}$ 					& \mr{2}{8.0}		& \mr{2}{1} 				& \mr{2}{ \citep{oos10} }		\\ 
							& \fns{(1000 - 1750 MHz)}		& $1.0\E{-30} \ergcmshz$					&  				&						& \vspace{0.04in}			\\ \hline
	\mr{2}{ASKAP} 				& Radio 					& $0.1  \textrm{ mJy}$ 					& 30.0			& \mr{2}{1} 				& \mr{2}{ \citep{joh09} }	 	\\ 
							& \fns{(700 - 1800 MHz)}		& $1.0\E{-27} \ergcmshz$					& (20,000)		&						& \vspace{0.04in}			\\ \hline
	\mr{2}{LOFAR}				& Low Radio 				& $1.0 \textrm{ mJy}$ 					& 3,000			& \mr{2}{1} 				& \mr{2}{ \citep{fen06} }	 	\\ 
							& \fns{(10 - 200 MHz)}		& $1.0\E{-26} \ergcmshz$					& (20,000)		&						& \vspace{0.04in}			\\ \hline
\end{tabular}
\footnotetext[1]{This sensitivity, as given in the literature, is specific for the 50 -- 300 \kev{} range.}
\caption[Characteristic Telescope and Survey Properties] {Characteristic telescope and survey properties used for analysis of electromagnetic transients associated with gravitational-wave progenitors.  These values are approximations to the true survey designs and strategies, which are, in some cases, yet to be determined.  Both the instantaneous Field of View (FoV), and the survey FoV --- corresponding to the listed cadences (cad.) --- are given in square-degrees.}
\label{table_instr}
\end{table*}

\subsection{Short GRBs}
\begin{figure*}[ht] 
	\centering
	\includegraphics[width=0.8\linewidth]{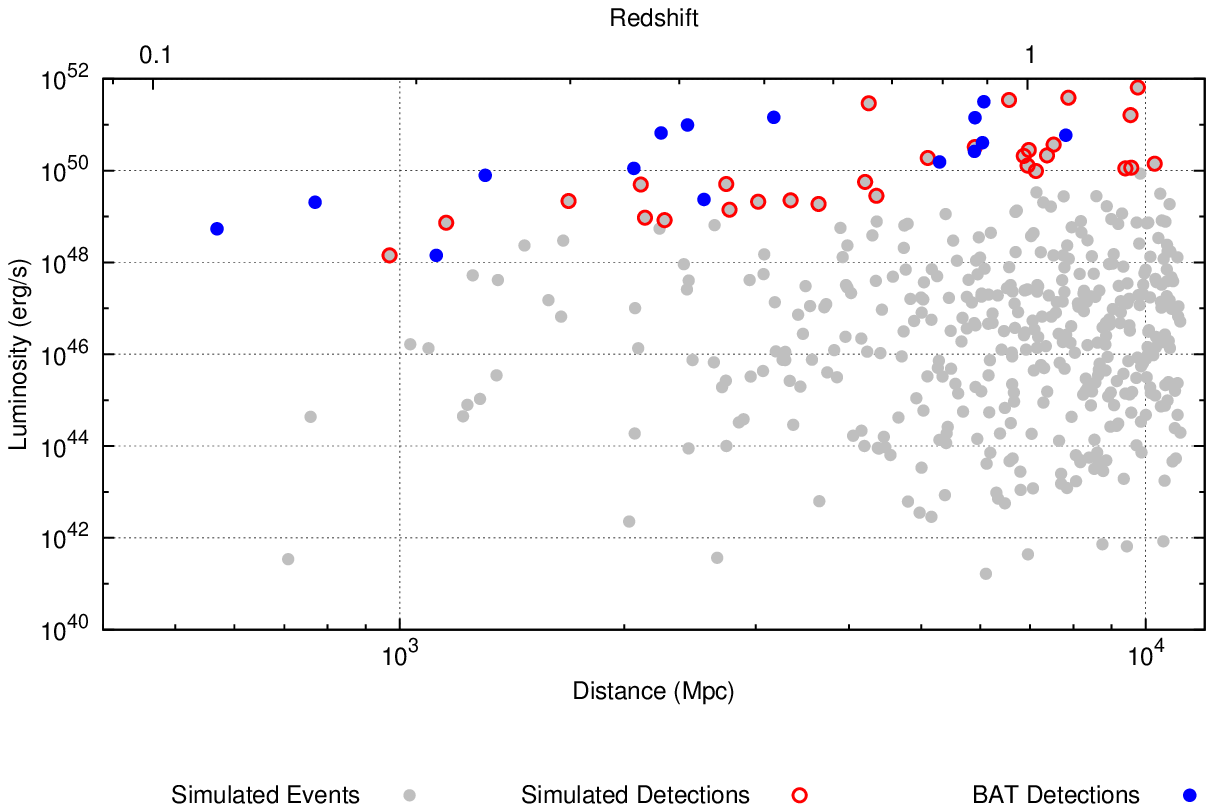}
	\caption[SGRB Luminosity Function]{SGRB luminosity and distance distribution for approximately a year of simulated events (grey) and simulated detections, with flux $\geq 10^{-8} \textrm{ erg cm}^{-2} \textrm{ s}^{-1}$ (red).  Isotropic equivalent luminosity in the BAT band is plotted against luminosity distance and redshift.  Over-plotted are 	the SGRBs with known distances from Table~\ref{table_tran_obs} (excluding GRB 090426 at $z=2.6$).  The simulated and observed SGRB properties appear consistent, and larger samples of data show populations with matching average properties.}
	\label{fig_lum_func}
\end{figure*}

From a sample of about 60 observed short bursts, roughly 16 have observed redshifts determined from spectroscopy of their associated host galaxy.  These bursts, compiled by \citet{ber10} are presented in Table \ref{table_tran_obs}.  In our simulation, the prompt emission in the BAT band is drawn from a log-normal luminosity distribution, constructed to roughly match the observed redshift--luminosity distribution of bursts, and the overall BAT event rate, for a flux selection cutoff of $10^{-8} \textrm{ erg cm}^{-2} \textrm{ s}^{-1}$.  The luminosity-function parameters are presented in Table~\ref{table_params}, and the distribution is compared with BAT detected luminosities in Fig.~\ref{fig_lum_func}.  The luminosity is extrapolated to the GBM band using the best-fit Band-model of BATSE data from \citet{kan06} (see Table~\ref{table_params}).  In both bands, the prompt emission is assumed to have a rest-frame duration of one second---approximately the average detected observer-frame value for the BAT.  The temporal uncertainty connecting SGRBs to GWs, however, is taken as six seconds as a more conservative upper limit, and consistent with the assumptions made in previous SGRB-triggered GW searches \cite[e.g.,][]{S6GRBLV}.  Finally, the emission is assumed to be constant within the jet angle, and zero outside.

\begin{table*}[ht] 
\renewcommand\arraystretch{1.0}
\centering      
\begin{tabular}{c c c c c c c }
\hpad{GRB}{0.1in}	& \hpad{z}{0.1in}	& Distance (Mpc)	& \hpad{$T_{90}$(s)}{0.01in}	& \hpad{$L_\msm{x}$ (erg/s)}{0.01in}	& \hspace{0.1in} \hpad{$T_\msm{opt}$ (hr)}{0.01in}		& \hpad{$L_\msm{opt}$ (erg/s/Hz)}{0.01in} 	\\ \hline \hline
050709			& 0.161			&	770			&	0.07					& $2.9\E{50}$						& \hspace{0.1in}	34.0							&	$1.6\E{27}$						\\ \hline
050724			& 0.257			&	1,302		&	3.00					& $2.6\E{49}$						& \hspace{0.1in}	12.0							&	$1.7\E{28}$						\\ \hline
051221A			& 0.546			&	3,172		&	1.40					& $1.0\E{51}$						& \hspace{0.1in}	3.1							&	$7.0\E{28}$						\\ \hline
061006			& 0.438			&	2,431		&	0.42					& $2.4\E{51}$						& \hspace{0.1in}	14.9							&	$2.1\E{28}$						\\ \hline
070714B			& 0.923			&	6,068		&	3.00					& $1.1\E{51}$						& \hspace{0.1in}	23.6							&	$3.1\E{28}$						\\ \hline
070724			& 0.457			&	2,558		&	0.40					& $5.9\E{49}$						& \hspace{0.1in}	2.3							&	$3.9\E{28}$						\\ \hline
071227			& 0.381			&	2,059		&	1.80					& $6.2\E{49}$						& \hspace{0.1in}	7.0							&	$8.1\E{27}$						\\ \hline
080905			& 0.122			&	568			&	1.00					& $5.4\E{48}$						& \hspace{0.1in}	8.5							&	$3.1\E{26}$						\\ \hline
090426			& 2.609			&	22,077		&	1.28					& $1.1\E{52}$						& \hspace{0.1in}	2.6							&	$1.2\E{31}$						\\ \hline
090510			& 0.903			&	5,905		&	0.30					& $4.7\E{51}$						& \hspace{0.1in}	9.0							&	$9.6\E{28}$						\\ \hline
100117			& 0.920			&	6044			&	0.30					& $1.4\E{51}$						& \hspace{0.1in}	8.4							&	$<1.3\E{28}$						\\ \hline
050509B			& 0.225			&	1,119		&	0.04					& $3.6\E{49}$						& \hspace{0.1in}	2.1							&	$<1.0\E{27}$						\\ \hline
060801			& 1.130			&	7,815		&	0.50					& $1.2\E{51}$						& \hspace{0.1in}	12.4							&	$<5.8\E{28}$						\\ \hline
061210			& 0.409			&	2,240		&	0.19					& $3.5\E{51}$						& \hspace{0.1in}	2.1							&	$<8.4\E{27}$						\\ \hline
061217			& 0.827			&	5,292		&	0.21					& $2.7\E{51}$						& \hspace{0.1in}	2.8							&	$<6.7\E{28}$						\\ \hline
070429B			& 0.902			&	5,896		&	0.50					& $5.2\E{50}$						& \hspace{0.1in}	4.8							&	$<2.5\E{28}$						\\ \hline \hline
Average			& 0.70			& 	4,707		&	0.90					& $5.9\E{49}$ 						& \hspace{0.1in}	11.7							&	$2.6\E{28}$						\\ \hline	
\end{tabular} 
\caption[Short GRB \xx-Ray and Optical Afterglow Properties] {Short GRB and optical afterglow properties from \citet{ber10}.  Redshifts have been converted to luminosity distances, and combined with the $T_{90}$ --- a typical measure of burst duration --- these were used to convert fluxes to isotropic-equivalent luminosity.  The `average' luminosities are medians, and do not include values with only upper limits.  All of the \xx-ray data (corresponding to $F_\gamma$ in \citet{ber10}) correspond to Swift BAT observations, except for GRBs 050709 and 060121.}
\label{table_tran_obs}
\end{table*} 

\subsection{Afterglows and r-Process Tidal Tails}
While short GRB prompt emission is constrained to the jet-axis, interaction of the relativistically jetted material with the circumburst medium produces strong emission at much broader angles, and is thus expected to be observable for an off-axis source as an `orphan afterglow.'  To model these afterglows, we use the synthetic afterglow library of \citet{van11}, with a jet energy of $10^{48}$ erg, an ambient density of $1 \textrm{ cm}^{-3}$ and a jet-angle of 0.2 radians; these values are presented in Table~\ref{table_params}.

\begin{table*}[ht] 
\renewcommand\arraystretch{1.0}
\centering      
\begin{tabular}{c c}
	\multicolumn{2}{c}{Transient Parameters}				 															\\ \hline 
	Jet Energy ($E_j$)									& $10^{48}$ erg											\\ 
	Circumburst Density ($n$)							& $1 \textrm{ cm}^{-3}$										\\ 
	SGRB Jet Half Opening-Angle ($\jetangle$)				& 0.2 rad													\\ 
	Band-Model GRB Spectra							& $\alpha = -1.08$, $\beta = -2.33$, $E_\textrm{peak} = 262$ KeV 		\\ 
	SGRB luminosity function									& \mr{2}{$\mu = 106.7$, $\sigma = 4.6$}						\vspace{-0.05in}	\\
	$p\left(\log \frac{L}{\textrm{erg s}^{-1}}\right) = \frac{1}{\sigma\sqrt{2\pi}} \exp\left({-\frac{[\log (L/(\textrm{erg\ s}^{-1}))-\mu]^2}{2\sigma^2}}\right)$	&													\\
	\multicolumn{2}{c}{Monte Carlo Parameters}																		\\ \hline 
	GW Horizon ($d_H$)								& 445 Mpc												\\ 
	GW Data Segment Duration ($\tau_\mty{GW}$)			& 0.1 s													\\ 
	Odds Ratio ($\Odds_\mty{EM} = \Odds_\mty{GW}$)			& $6\E{15}$												\\  
	Merger Rate ($\mathcal{R}$)							& $1 \textrm{ Mpc}^{-3} \textrm{ Myr}^{-1}$						\\ 
	\mr{2}{Cosmological Parameters\fnm{1}}					& $\Omega_\Lambda = 0.734$, $\Omega_b = 0.0449$, $\Omega_c = 0.222$ \vspace{-0.05in} \\
													& $H_0 = 71.0 \textrm{ km s}^{-1} \textrm{ Mpc}^{-1}$ 		\\ \hline
\end{tabular} 
\footnotetext[1]{\citet{wmap7}}
\caption[Calculation Parameters] {Afterglow simulation parameters from \citet{van11}, best-fit Band model parameters from \cite{kan06}, and additional values used in the current study.}
\label{table_params}
\end{table*} 

\begin{figure*}[ht] 
	\centering
	\includegraphics[width=0.9\linewidth]{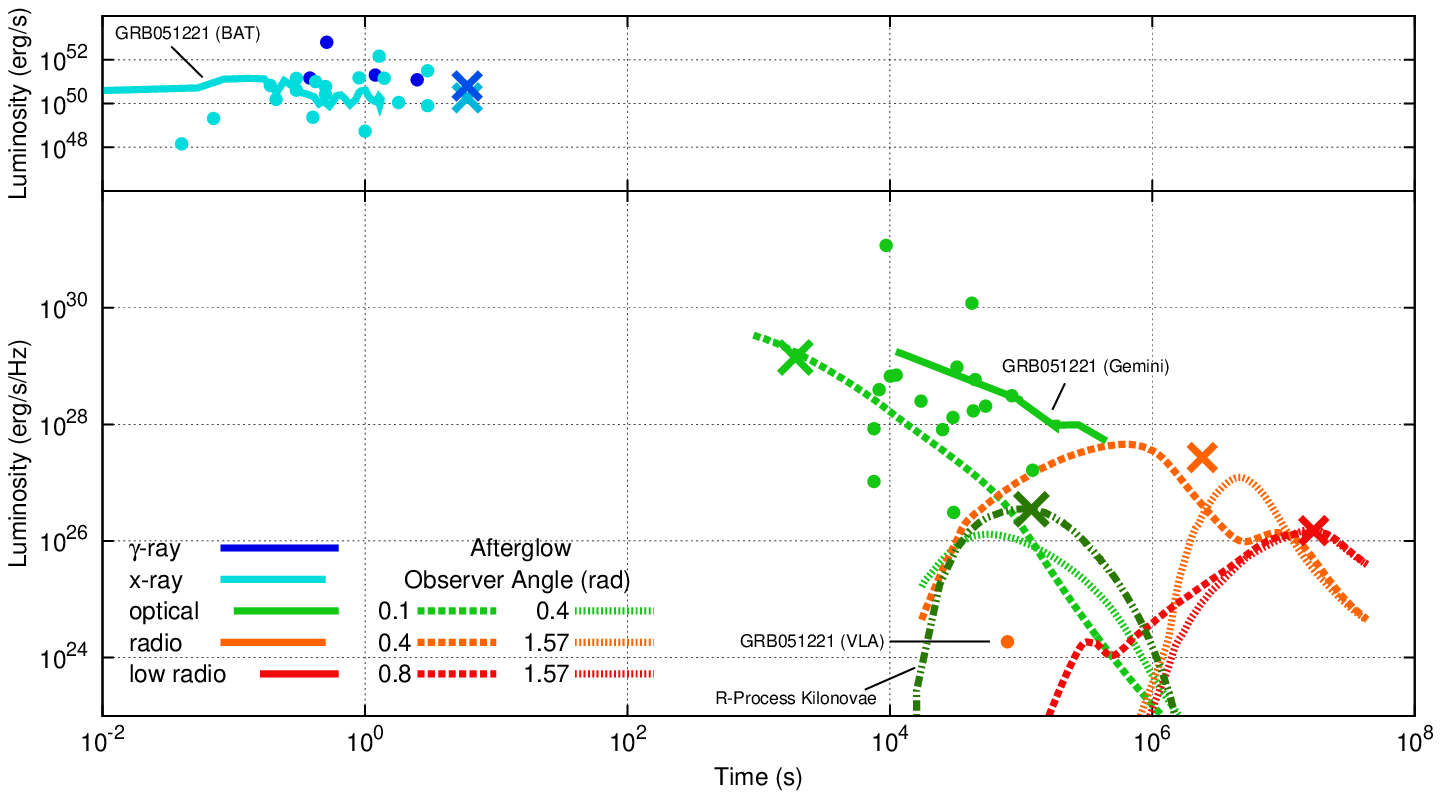}
	\caption[Electromagnetic Transients]{Electromagnetic Transients Associated with Compact Binary Mergers.    Dots denote peak luminosities of observed SGRBs and afterglows.   For illustration, SGRB 051221 BAT and optical (Gemini-N/GMOS) light curves from \citet{sod06} are plotted as solid lines, with a single radio (VLA) afterglow observation as an orange dot.  Simulated afterglows and r-process tidal tail emission (dark green) are shown with dashed lines for different observer angles, as described in the legend.  Simulated isotropic-equivalent average luminosities and TaMs are plotted with `$\times$'s.}
	\label{fig_transients}
\end{figure*}

R-process powered emission from the tidally ejected NS tails is expected to have comparable luminosities to optical afterglows, on shorter timescales for typical viewing angles.  In our analysis we use the lightcurves generated by \citet{rob11}, in their BNS model.  Figure \ref{fig_transients} shows the characteristic light-curves for each type of transient used in our simulation, along with several observed transients for comparison.  It is possible that opacity for r-process events is much larger than expected \cite{kase13, barn13}, which would decrease their luminosity and drive the main emitted energy to longer wavelengths.  Thus, our assumptions here may prove optimistic.

\section{Results}
The framework outlined above can be used to estimate typical values of the prior and \z-factor---the increase in sensitivity to GW events when associated with an EM transient.  Consider the detection of a SGRB, for example, which could be observed within roughly 10 Gpc.  If we assume that LIGO could detect an associated GW signal out to about 200 Mpc, we can use the parameters listed in Tables~\ref{table_eta} \& \ref{table_params} in Eq.~\ref{eq_snr_ratio} to obtain a modest improvement of $\zeta \approx 1.07$.  The increase in volume to which LIGO would be sensitive to GWs, with a SGRB detection, is then roughly 20\%.  As will be seen later, however, this does \textit{not} equate to a proportional increase in the overall rate of GW detections ---  because most GW detections will not be associated with observable SGRBs.  These estimates are consistent with the results of simulations, presented below.

The average properties of Monte Carlo events which satisfied their telescope's detection criteria are presented in Table~\ref{table_em_props}.  Observed and simulated SGRBs and afterglows are plotted in Fig.~\ref{fig_transients}.  In general, the simulated detection properties are consistent with the distance and luminosity distribution of observed SGRBs (Table~\ref{table_tran_obs}), as reinforced in Fig.~\ref{fig_lum_func}.  However, note that we have assumed a time-at-maximum (TaM) of 6 seconds for short GRBs for consistency with the wider time window used in previous GRB-triggered GW searches \cite{S5GRBLV,S6GRBLV} to allow for a margin of safety.  We also find that the average simulated TaM for optical afterglows is shorter than for SGRB afterglow observations.

The average fraction by which the SNR detection threshold is lowered for a given candidate GW signal associated with an EM transient---the \z-factor---is listed for each transient-telescope combination in Table \ref{table_em_zeta}.  Values are given for both the events which do, and do not, have redshift determination.  Note that \z~is the change in detection threshold for a given event---the average of which is presented in Table~\ref{table_em_zeta}.  These mean values are averaged over all detected EM events in the simulation---not only those which lead to triggered GW-detections, which are most significantly aided by transients with redshift determination.  The vast majority of events from instruments with large detector horizons (e.g.,~BAT and afterglows with LSST) will occur outside of the LIGO-Virgo sensitive volume, which has a BNS horizon distance $d_H \sim 445$ Mpc. Similarly, instruments with detector horizons significantly less than the GW horizon (e.g.,~LOFAR and ASKAP) will not increase the number of GW detections regardless of the value of \z, because any event within their detectable volume will already exceed the blind-detection SNR threshold.  

\begin{table*}[ht] 
\renewcommand\arraystretch{1.1}
\centering      
\begin{tabular}{ c  c  c  c  c  c  c }
								&				& \multicolumn{5}{c}{Mean EM Properties}														\\ \cline{3-7}                           
	\hpad{Transient}{0.05in}			& Telescope \,		& $d_L$ \fns{(Mpc) }	& Lum$\ergshz$		& Angle {(rad)}	& ToT {(s)}		& \hpad{TaM {(s)}}{0.1in}		\\ \hline \hline
	\multirow{2}{*}{SGRB \fns{Prompt }}	& BAT			& 5200			& $1.8\E{50} \ergs$		& 0.13		& -				& -						\\ 
								& GBM			& 4800			& $6.3\E{50} \ergs$		& 0.13		& -				& -						\\ \cline{1-7} 
	\multirow{5}{*}{Afterglow}			& LSST			& 4100			& $1.4\E{29}$			& 0.16		& $4.0\E{4}$		& $1.9\E{3}$				\\ 
								& PTF			& 1600			& $2.4\E{29}$			& 0.13		& $9.1\E{3}$		& $1.3\E{3}$				\\ \cline{2-7}
								& Apertif			& 3200			& $2.7\E{27}$			& 0.82		& $3.6\E{6}$		& $2.4\E{6}$				\\ 
								& ASKAP			& 130			& $3.3\E{27}$			& 0.69		& $1.7\E{6}$		& $1.4\E{6}$				\\ \cline{2-7}
								& LOFAR			& 8.26			& $1.5\E{26}$			& 1.0			& $2.7\E{7}$		& $1.7\E{7}$				\\ \hline \hline
	\multirow{2}{*}{R-Process}		& LSST			& 460			& $3.5\E{26}$			& 1.0			& $2.4\E{5}$		& $1.2\E{5}$				\\ 
								& PTF			& 92				& $3.5\E{26}$			& 1.0			& $2.4\E{5}$		& $1.1\E{5}$				\\ \hline \hline

\end{tabular} 
\caption[Average Electromagnetic Properties]{Average properties of simulated electromagnetic transients.  Each property is the average over detected events for the given telescope.  The time-over-threshold (ToT) and time-at-maximum (TaM) are calculated in the observer frame, and the luminosity is the average peak, isotropic-equivalent.  } 
\label{table_em_props}
\end{table*}

The value of \z~is increased by a higher prior on the presence of a GW signal, and by better constraints on the viable GW parameter space.  The best improvement to the prior comes from an EM-sensitive volume comparable to (or smaller than) the GW-volume, and from the most precise temporal accuracy.  In our model, temporal localizability is determined entirely by the time-at-maximum  (TaM), while the accuracy in position space is relatively constant (see Sec.~\ref{sec_parspace}), unless a redshift (distance) can be determined.  The TaM approximation is motivated by the uncertainty which would be associated with making only a single observation of a given transient, while having accurate models of its source.  Positional accuracy, without redshift, is limited by the precision of GW triangulation (significantly worse than any EM localization), and thus constant between transient types.  Our requirement for redshift determination is based on a time-over-threshold (ToT) longer than a day.  Based on the ToT values in Table~\ref{table_em_props}, redshift determination is rare in optical afterglows, and never occurs for SGRB prompt emission alone.

\begin{table*}[ht] 
\renewcommand\arraystretch{1.2}
\centering      
\begin{tabular}{ c  c  c  c  c }
								&						& \multicolumn{3}{c}{Mean \z}																				\\ \cline{3-5} 
	\hpad{Transient Type}{0.1in}		& \hpad{Telescope}{0.2in}	& \hpad{With Redshift}{0.0in} 			& \hpad{Without Redshift}{0.0in} 		& \hpad{Overall}{0.0in}		 			\\ \hline \hline
	\multirow{2}{*}{SGRB Prompt}		& Swift (BAT)				& -								& 1.063							& 1.063								\\ \cline{2-5}
								& Fermi (GBM)				& -								& 1.063							& 1.063								\\ \hline
	\multirow{5}{*}{Afterglow}			& LSST					& 1.131							& 1.000							& 1.014								\\ \cline{2-5}
								& PTF					& 1.135							& 1.035							& 1.035								\\ \cline{2-5}
								& Apertif					& 1.053							& 1.000							& 1.015								\\ \cline{2-5}
								& ASKAP					& 1.058							& 1.015							& 1.025								\\ \cline{2-5}
								& LOFAR					& 1.034							& 1.000							& 1.007								\\ \hline \hline
	\multirow{2}{*}{R-Process Tails}	& LSST					& 1.084							& 1.038							& 1.056								\\ \cline{2-5}
								& PTF					& 1.085							& 1.038							& 1.058								\\ \hline \hline
\end{tabular} 
\caption[Zeta values.]{Lowered detection threshold factor \z.  If the criteria for redshift determination are satisfied, we assume the distance to the event is fully determined, and the \z-factor is increased.  The mean value of \z~reflects the fraction of simulated EM detections with redshift determination.}
\label{table_em_zeta}
\end{table*} 

As mentioned earlier, the value of \z~alone does not fully determine the benefits of triggered searches.  It must be considered together with the sensitivity of a given electromagnetic search and the total number of transients expected to be observable with it.  For example, the highest average value of \z~occurs for PTF observations of r-process kilonovae; however, the typical distance for such events is under 100 Mpc---and thus any events within that volume are already likely to exceed the blind GW-search threshold.  Additionally, the expected rate of kilonovae detections with PTF, within that volume, is less than one per year.  The expected EM transient rates are presented in Table~\ref{table_em_rates}.  It is apparent that the expected detection rates by LOFAR for afterglows, and PTF for those and kilonovae, are too small for much benefit.

\begin{table*}[ht] 
\renewcommand\arraystretch{1.2}
\centering      
\begin{tabular}{ c  c  c  c  }
	\hpad{Transient Type}{0.1in}		& \hpad{Emission}{0.2in}	& \hpad{Telescope}{0.2in}	& \hpad{EM Rate (yr$^{-1}$)}{0.1in} 				\\ \hline \hline
	
	\multirow{2}{*}{SGRB Prompt}		& \xx-ray				& Swift (BAT)				& 29.1		$\pm$	0.4			\\ \cline{2-4}
								& Gamma/\xx-ray		& Fermi (GBM)				& 71.2		$\pm$	0.3			\\ \cline{1-4}
	\multirow{5}{*}{Afterglow}			& \multirow{2}{*}{Optical}	& LSST					& 69			$\pm$	2			\\ \cline{3-4}
								& 			 		& PTF					& 1.1 		$\pm$	0.2			\\ \cline{2-4}
								& \multirow{2}{*}{Radio}	& Apertif					& 8			$\pm$	1			\\ \cline{3-4}
								& 			 		& ASKAP					& 5.80		$\pm$	0.01			\\ \cline{2-4}
								& Low Radio			& LOFAR					& $2.678\E{-3}$ $\pm$ $3\E{-6}$		\\ \hline \hline
	\multirow{2}{*}{R-Process Tails}	& \multirow{2}{*}{Optical}	& LSST					& 146.0 		$\pm$	0.1			\\ \cline{3-4}
								&					& PTF					& 0.880		$\pm$	0.002		\\ \hline
\end{tabular} 
\caption[Electromagnetic Detection Rates]{Rates of electromagnetic transient detections produced by our simulations.  The intrinsic merger-rate density is taken as $10^{-6} \textrm{ Mpc}^{-3} \textrm{ yr}^{-1}$, in comoving volume and the source's rest-frame, from \citet{ratesdoc}.} 
\label{table_em_rates}
\end{table*} 

The expected rates of gravitational-wave detections from EM triggered searches are presented in Table~\ref{table_gw_rates}, along with the factor increase (gain) relative to the `blind' (non-triggered) detection rate of about 30.8 yr$^{-1}$.  Observations of kilonovae with a deep, wide survey like LSST suggest that the GW detection rate could be boosted by almost 10\%, while all other transients show gains at a one-percent level or below.  The rarity of possible coincident detections underlies the negligible enhancement to the GW detection rate triggered by SGRBs.  In particular, for SGRBs, $\zeta \approx 1.06$, thus the GW sensitive volume is increased by roughly a factor of $\zeta^3 \approx 1.2$, and the SGRB+GW detection rate increases from $\sim 0.09$ to $\sim 0.1$---which, when compared to a baseline detection rate of 32, is a $< 0.1\%$ increase, as seen in Table~\ref{table_gw_rates}. These results suggest that triggered searches offer small or effectively negligible gains to the rate of gravitational wave detections alone.

\begin{table*}[ht] 
\renewcommand\arraystretch{1.2}
\centering      
\begin{tabular}{ c  c  c  c  }
								&						& \multicolumn{2}{c}{GW Detections (yr$^{-1}$)}			\\ \cline{3-4}                           
	Transient Type					& Telescope				& \hpad{Triggered}{0.2in}		& Gain vs.~blind
	\\ \hline \hline
	\multirow{2}{*}{SGRB Prompt}		& Swift (BAT)				& 0.01 $\pm$ 0.003	 		& 1.000 $\pm$ 0.001		\\ \cline{2-4}
								& Fermi (GBM)				& 0.06 $\pm$ 0.02	 		& 1.002 $\pm$ $1\E{-3}$		\\ \cline{1-4} 
	\multirow{5}{*}{Afterglow}			& LSST					& 0.51 $\pm$ 0.09			& 1.017 $\pm$ 0.004		\\ \cline{3-4}
								& PTF					& 0.00 $\pm$ 0.01	 		& 1.000 $\pm$ 0.006		\\ \cline{2-4}
								& Apertif					& 0.00 $\pm$ 0.04			& 1.000 $\pm$ 0.002		\\ \cline{3-4}
								& ASKAP					& 0.13 $\pm$ 0.07 			& 1.004 $\pm$ 0.003		\\ \cline{2-4}
								& LOFAR					& 0.00 $\pm$ 0.07	 		& 1.000 $\pm$ 0.003		\\ \hline \hline							
	\multirow{2}{*}{R-Process Tails}	& LSST					& 2.74 $\pm$ 0.02		 	& 1.089 $\pm$ $1\E{-3}$		\\ \cline{3-4}
								& PTF					& 0.03 $\pm$ 0.02	 		& 1.001 $\pm$ 0.001		\\ \hline \hline
\end{tabular} 
\caption[Gravitational-Wave Detection Rates]{Rate of gravitational-wave detections from an EM-triggered search, and gain factor relative to the blind, all-sky GW detection rate of $\sim 30.8$ detections per year for the chosen merger rate.   The presence of certain electromagnetic transients increases the prior probability of LIGO data containing a detectable signal.  Furthermore, information from the EM observation better constrains the allowed GW parameter space.   Statistical uncertainties from Monte Carlo modeling are included.} 
\label{table_gw_rates}
\end{table*} 

On the other hand, EM-triggered GW searches could give a significant relative boost to the number of {\it multi-messenger} observations. Assuming $10^{-6}$ BNS mergers per Mpc$^3$ of comoving volume per year, advanced detectors are expected to make tens of GW detections a year.  The intrinsic rate of GW and electromagnetic coincident detections, however, is much lower --- and shows a much more noticeable improvement from triggered searches.  The rate enhancement for coincident detections is presented in Table~\ref{table_gwem_rates}.  We compare blind searches (in which simultaneous detections of EM and GW transients from the same event are purely fortuitous) with a combination of blind and EM-triggered searches.  Triggers made by high-energy observations of SGRBs show improvements of about 15\%.  Still, with both Swift and Fermi, such coincident detections would only be expected once every two to ten years.  

LSST observations of both afterglows and kilonovae show a noticeable rate enhancement of $\sim 30\%$ and $\sim 40\%$, respectively.  In the afterglow case, the boosted rate is still just over two coincident events per year for the default BNS merger rate, while that of kilonovae increases from roughly 7 to 10 detections per year.

\begin{table*}[ht] 
\renewcommand\arraystretch{1.2}
\centering      
\begin{tabular}{ c  c  c  c  c  }
								&						& \multicolumn{3}{c}{GW+EM Coincident Rate}									\\ \cline{3-5}                           
	Transient Type					& Telescope				& \hpad{Blind}{0.4in}		& \hpad{Blind + Triggered}{0.2in}	& Gain				\\ \hline \hline
	\multirow{2}{*}{SGRB Prompt}		& Swift (BAT)				& 0.0908 $\pm$ $9\E{-4}$		& 0.1046 $\pm$ $9\E{-4}$			& 1.15 $\pm$ 0.02		\\ \cline{2-5}
								& Fermi (GBM)				& 0.368 $\pm$ 0.002 		& 0.422 $\pm$ 0.002 			& 1.147 $\pm$ 0.008	\\ \cline{1-5} 		
	\multirow{5}{*}{Afterglow}			& LSST					& 1.80 $\pm$ 0.02	 		& 2.31 $\pm$ 0.02				& 1.28 $\pm$ 0.02		\\ \cline{3-5}
								& PTF					& 0.096 $\pm$ 0.005 		& 1.008 $\pm$ 0.005			& 1.05 $\pm$ 0.07		\\ \cline{2-5}
								& Apertif					& 0.0051 $\pm$ $6\E{-4}$		& 0.007 $\pm$ $0.001$		 	& 1.3 $\pm$ 0.2		\\ \cline{3-5}
								& ASKAP					& 4.64 $\pm$ 0.05			& 4.77 $\pm$ 0.05 				& 1.03 $\pm$ 0.02		\\ \cline{2-5}
								& LOFAR					& 0.0028 $\pm$ $8\E{-4}$		& 0.0028 $\pm$ $8\E{-4}$			& 1.0 $\pm$ 0.4		\\ \hline \hline
	\multirow{2}{*}{R-Process Tails}	& LSST					& 7.14 $\pm$ 0.02	 		& 9.88 $\pm$ 0.03	 			& 1.384 $\pm$ 0.005	\\ \cline{3-5}
								& PTF					& 0.661 $\pm$ 0.001 		& 0.69 $\pm$ $0.001$			& 1.044 $\pm$ 0.002	\\ \hline \hline
\end{tabular} 
\caption[Coincident Detection Rate of Both GW and EM Signals]{The coincident gravitational-wave and electromagnetic detection rates, for both blind-search detections made serendipitously, and detections made with a combination of a blind search and searches triggered on EM-transient observations.  Due to the lower intrinsic probability of making blind coincident detections, the rate enhancement from triggered searches is much more significant for coincident detections than for GW detections alone.} 
\label{table_gwem_rates}
\end{table*}

\newpage

\section{Discussion}
\label{sec:disc}
In this study, we have outlined a Bayesian framework for evaluating the increased observational sensitivity of gravitational-wave detectors in searches triggered by electromagnetic transients.  To determine the plausible benefits of such triggered searches, we apply this framework to a variety of electromagnetic transients associated with binary neutron-star mergers, and a series of telescopes and surveys to identify them.  These triggered searches decrease the required signal-to-noise threshold for a positive detection through the greater \textit{a priori} probability of the presence of a GW signature in the associated LIGO/Virgo data and tighter constraints on the parameter space of possible signals.  We find that observations of r-process kilonovae by a deep, wide-field transient survey like LSST provides the maximum benefits --- increasing the rate of multimessenger detections by almost 40\%. Once advanced LIGO/Virgo and LSST are at design specifications, we predict that using such triggered gravitational-wave searches could increase the detection rate to about 10 coincident detections per year.  Using optical triggers from both kilonovae and SGRB orphan afterglows, along with high-energy triggers from Swift and Fermi could together increase the coincident detection rate to about 14 per year, although some of the binary mergers may be double-counted as, e.g., Swift/Fermi and LSST triggers.

This type of multi-messenger astronomy offers tantalizing prospects for probing ultra-compact objects, their binary dynamics, and their eventual merger; in addition to possible tests of cosmology in the low-redshift universe, and possible insights into the origin of r-process nucleosynthetic elements.  A SGRB--GW coincident detection might be the most exciting prospect, as it would represent a definitive determination of the progenitor to these energetic outbursts.  In the explicit absence of such a coincidence, the use of triggered searches also improves the range at which such a detection could exclude the binary progenitor hypothesis \cite{LIGOGRB}.

While SGRB triggers offer the highest timing accuracy, the uncertainty in their distance and the low event rate within the LIGO sensitive volume make them a sub-optimal trigger with current \xx-ray instruments.  Keeping in mind that we have assumed a very simple luminosity function (Fig.~\ref{fig_lum_func}), it suggests that as many as half of SGRBs within a few hundred megaparsecs could be undetectable, even if jetted towards the earth.  An \xx-ray telescope with the field-of-view of the GBM and a sensitivity an order of magnitude higher than the BAT might significantly boost the rate of detections within the LIGO-volume, and thus the benefits to triggered GW searches.

A factor of ten to one hundred more optical kilonovae than SGRBs could be observed from within the GW-detection volume (depending on the assumed SGRB beaming angle).  It is important to note that the exact peak times and peak temperatures of kilonovae depend on the currently unconstrained line opacities of r-process elements.  Under the assumptions used here, these kilonovae observations could significantly enhance the rate of multimessenger detections, despite their low timing precision relative to SGRBs.  Surveys like PTF---with a sensitivity to about 21st magnitude---aren't able to probe deeply enough to boost the LIGO threshold.   Based on our findings, the most productive electromagnetic survey to trigger gravitational-wave searches would be an optical survey with comparable cadence and sky-coverage to PTF, but which is about an order of magnitude more sensitive (i.e.~reaching about 23rd magnitude).  LSST, with a sensitivity of about 24th magnitude, is more than sufficient to fully capture events in the LIGO sensitive volume {\it within the LSST survey field of view}---and optimize the rate of multi-messenger observations.   While LSST triggers increase the multi-messenger detection rate by almost $40\%$ (under the assumption that kilonovae release most of their energies at optical energies, which is uncertain), the boost to the overall GW detection rate is just under $10\%$ because LSST only observes about one-quarter of the sky.

Searches triggered on EM transients will add the same number of additional detections to both the total of GW detections and the number of coincident GW+EM detections.  However, the fractional gain in the GW detection rate will be smaller than the gain in the rate of coincident detections by a factor equal to the sky coverage of the relevant instrument over the timescale relevant to the transient being observed.  Our results suggest that despite the promising prospects for coincident detections, the use of triggered searches would only marginally increase the total number of detected gravitational-wave signals, regardless of survey strategy.

The pioneering study of triggered GW-searches, carried out by \citet{koc93}, predicted a rate enhancement of about a factor of 3---significantly larger than that found in the current work.  The discrepancy between these results is due to the difficulty of extracting a GW signal from noisy data.  An underlying assumption of the \citet{koc93} analysis is that the detector noise is both Gaussian and stationary; whereas in practice the noise can be correlated and `glitchy', which then requires a higher detection threshold to achieve the same false alarm rate.  Additionally, multiple filter templates are required for a search in which the component masses are not known, constituting a so-called `trials factor', which increases the false alarm rate for a fixed SNR threshold.  Finally, \citet{koc93} assume a one-to-one correspondence between SGRBs and GW signals, which allows them to ignore data corresponding to times without a SGRB observation. 

In our analysis, we have made specific assumptions regarding the necessary odds ratio for confident detection in the presence of glitchy noise, the expected sensitivity of advanced LIGO and Virgo, and the binary merger rate.  It is worth considering how changes to these assumptions might affect our results.  As illustrated by the disagreement between our results and those of \citet{koc93}, should future detectors achieve better-behaved data (improved data quality for a fixed average noise spectrum), the improvement from triggered searches would be enhanced.  The best improvements from triggered searches come from EM telescopes with sensitive volumes comparable to those of the GW detectors.  Therefore, if the GW-detector noise spectrum were lowered over time, increasing the LIGO-Virgo sensitivity, deeper EM surveys would become useful.  Perhaps the most uncertain parameter is the BNS merger rate.  If the binary merger rate were lower than expected, the prior GW probability $p({\textrm GW})$ would decrease while the conditional probability given an EM transient detection, $p({\textrm GW}|{\textrm EM})$, would remain the same.  Thus, while a lower intrinsic merger rate would decrease the overall rate of GW detections, the benefit and importance of triggering would be enhanced.

In this paper, we have attempted to more precisely determine the plausible benefits of triggering gravitational-wave searches on electromagnetic transients.  At the same time, the statistical framework we have formulated for analyzing the expected enhancement from using multiple observational channels is completely generalizable to any system of correlated observations.  The same technique could easily be applied to space-based interferometers, such as LISA \cite{LISA,Danzmann:1997} or NGO \cite{NGO}; or be used as a boost to high-energy particle astronomy --- such as with Veritas \cite{veritas}, or neutrino astronomy --- with, e.g., IceCube \cite{icecube}.

\begin{acknowledgments}
We are grateful to Hendrik van Eerten for making afterglow simulation data available online and for providing us with additional simulations for this project, and to Luis Lehner, Edo Berger, and especially Peter Shawhan for comments on the manuscript. 
\end{acknowledgments}

\newpage
\bibliography{EMtriggers}

\end{document}